%% file: paper.tex
\documentclass[12pt]{article}
\usepackage{epsfig}  \usepackage{graphics}
\usepackage{my}
\newcommand{\dg}{\ensuremath{^{\circ}}}
\begin{document} 
\title{Small-$x$ Physics and Forward Jet Production at THERA}
\author{Hannes Jung$^1$ and Leif L\"onnblad$^2$}
\maketitle
{\small $^1$ Department of Physics, Lund University, Sweden, jung@mail.desy.de

{\small $^2$ Department of Theoretical Physics, Lund University, Sweden, 
leif@thep.lu.se}
\begin{abstract}
  We discuss some aspects of forward jet production as a signature for small $x$
  physics at THERA energies.
\end{abstract}
\newcommand{\CCFM}{CCFMa,CCFMb,CCFMc,CCFMd}
\newcommand{\DGLAP}{DGLAPa,DGLAPb,DGLAPc,DGLAPd}
\newcommand{\BFKL}{BFKLa,BFKLb,BFKLc}
\newcommand{\asb}{{\bar \alpha}_\mathrm{s}}
\newcommand{\as}{\alpha_\mathrm{s}}
\def\CASCADE{{\sc Cascade}}
\def\SMALLX{{\sc Smallx}}
\def\RAPGAP{{\sc Rapgap}}
\def\LDC{{\sc Ldc}}
\section{Introduction}

The evolution of the parton densities at small $x$ is a very rich but
complicated issue. The steep rise of the structure function $F_2$ at
small $x$ is explained by the presence of a huge gluon number density.
The pure DGLAP~\cite{\DGLAP} evolution equations, meant to describe
the evolution of the parton densities as a function of $Q^2$, are able
to reproduce the rise of $F_2$ provided the input starting
distributions are chosen properly.
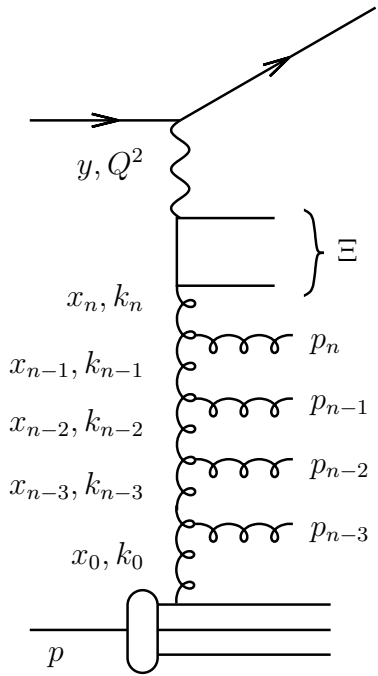
\begin{figure}[htb]
\begin{center} 
\input{ladder.pstex_t}
\end{center} 
\caption{{\it Kinematic variables for multi-gluon emission. The
    $t$-channel gluon momenta are given by $k_i$ and the gluons
    emitted in the initial state cascade have momenta $p_i$. The upper
    angle for any emission is obtained from the quark box, as
    indicated with $\Xi$.
\label{CCFM_variables} }} 
\end{figure}
\par
Figure~\ref{CCFM_variables} shows the pattern of QCD initial-state
radiation in a small-$x$ DIS event, together with labels for the
kinematics. The gluon splitting function $P_{gg}$ is given by:
\begin{equation}
P_{gg}(z_i,k^2) = \asb \left( \frac{1}{z_i} -2 + z_i(1-z_i) +
\frac{1}{1-z_i} \right) 
\frac{1}{k^2},
\label{Pgg}
\end{equation}
where $\asb = \as C_A/\pi$ and $z_i= x_i/x_{i-1}$ (see
Fig.~\ref{CCFM_variables}) is the ratio of the energy fractions of
successive branchings in the gluon chain and $k^2$ is virtuality of
the $t$-channel gluon with $k^2 \sim k_t^2$. In DGLAP the $k^2$
dependence of all emissions in the gluon chain are simplified by the
observation that at not too small $x$ the dominant part to the cross
section comes from the region of phase space where $k^2$ is very
small. However if $x$ or $z$ becomes very small, the collinear (small
$k_t^2$) approximation of DGLAP may be inadequate.  This region is
treated by the BFKL~\cite{\BFKL} evolution equation, which keeps the
full $k_t^2$ integration but approximates the gluon splitting function
with the asymptotic form $$P_{gg}\sim \frac{1}{z} \frac{1}{k_t^2}. $$
\par
From a detailed analysis of interference effects in a gluon chain, it
was found~\cite{\CCFM} that the proper evolution variable is the angle
of the emitted gluon, and not the virtuality $k^2$ as in the DGLAP
approximation, nor $z$ as in the BFKL approximation. This angular
ordering resulted in the new, and more complicated, CCFM
evolution~\cite{\CCFM}, which reproduces the BFKL and DGLAP
approximations in the small and large $x$ limits respectively. The
CCFM equation naturally interpolates between the two extremes.
However, in CCFM the gluon splitting function contains only the
singular terms in $z$:
$$P_{gg} = \asb\left(\frac{1}{z}\Delta_{ns}+\frac{1}{1-z}\right),$$
with $\Delta_{ns}$ being the non-Sudakov form factor to regulate the
$1/z$ singularity. The non singular terms of the splitting function
(see eq.(\ref{Pgg})) are not obtained within the CCFM approximation.
\par
Whereas at HERA energies ($\sqrt{s} \sim 300$ GeV) the total cross
section of deep inelastic scattering can be reasonably well described
with the DGLAP evolution equations, measurements of specific features
of the hadronic final state indicate clear deviations from a pure
DGLAP scenario.
\par
The cross section at low $x$ and large $Q^2$ for a high $E^2_T$ jet in
the proton direction (a forward jet) has been advocated as a
particularly sensitive measure of small $x$ parton dynamics
\cite{Mueller_fjets1,Mueller_fjets2}. If the forward jet has large
energy ($x_{jet}=E_{jet}/E_{proton} \gg x$) the evolution from
$x_{jet}$ to small $x$ can be studied.  When $E_T^2 \sim Q^2$ there is
no room for $Q^2$ evolution left and the DGLAP formalism predicts a
rather small cross section in contrast to the BFKL/CCFM formalisms,
which describe the evolution also in $x$. Measurements performed at
HERA~\cite{H1_fjets_data,ZEUS_fjets_data} show that the prediction
from the naive DGLAP formalism lies a factor $\sim 2$ below the data,
whereas the data can be described by CCFM evolution
equations~\cite{jung_salam_2000}.

\section{Initial State QCD Cascade}
\label{sec:cascade}
The effect of new small $x$ parton dynamics is most clearly seen if the
contribution from typical DGLAP dynamics is suppressed. The forward jet
production in deep inelastic scattering at small values of $x$ is one example of
such a process. However, the kinematics need to be investigated further. Typical
event selection criteria at HERA are:
\begin{center}
\begin{tabular}{l}
$Q^2>10$ GeV$^2$ \\
$E_{t\;jet} > 5$ GeV \\
$\eta_{jet} < 2.6$ \\
$x_{jet}>0.036$\\ 
$0.5<E^2_T/Q^2<2$
\end{tabular} 
\end{center}
\begin{figure}[htb]
  \begin{center}
  \epsfig{file=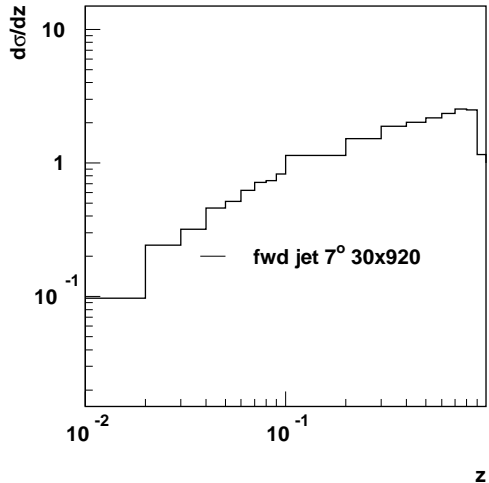,height=8.cm}
  \caption{{\it The values of the splitting variable $z$ for events satisfying the
  forward jet criteria, with $\theta = 7\dg$ at HERA energies. }
}
  \label{zval_fjet_hera} 
\end{center}
\end{figure}
The range in $x$ is typically $10^{-3} < x < 10^{-2}$. The evolution
takes place from the large $x_{jet}$ down to the small $x$ with a
typical range at HERA energies of $\Delta x=x/x_{jet}\sim 0.1 - 0.01$.
In order to justify the use of an evolution equation (instead of a
fixed order calculation) one would require at least 2 or more gluon
emissions during the evolution. To roughly estimate the energy
fractions $z_i$ of 3 gluon emissions between $ 10^{-3} < x < 10^{-1}$,
one can assume that each gluon carries the same energy. Then the range
of $\Delta x \sim 0.01$ results in $z \sim 0.2 $, which is far from
being in the very small $z$ region, where the BFKL or CCFM
approximations (treating only the $1/z$ terms in the gluon splitting
function) are expected to be appropriate. In Fig.~\ref{zval_fjet_hera}
we show the values of the splitting variable $z$ in events satisfying
the forward jet criteria at HERA energies obtained from the Monte
Carlo generator \CASCADE~\cite{jung_salam_2000}.  Since the values of
the splitting variable $z$ are indeed in the large $z$ region (the
majority has $z>0.1$), it is questionable, whether the BFKL or CCFM
evolution equations, including only the $1/z$ terms of the gluon
splitting function, are already applicable.  Whereas the measurement
at HERA stops at $x\sim 10^{-3}$, the available phase space at THERA
is enlarged by a factor of $\sim10$. Therefore the gluons along the
chain will presumably have smaller $z$ values and the usage of the
small $x$ evolution equations might be more justified.
\begin{figure}[htb]
  \begin{center}
  \epsfig{file=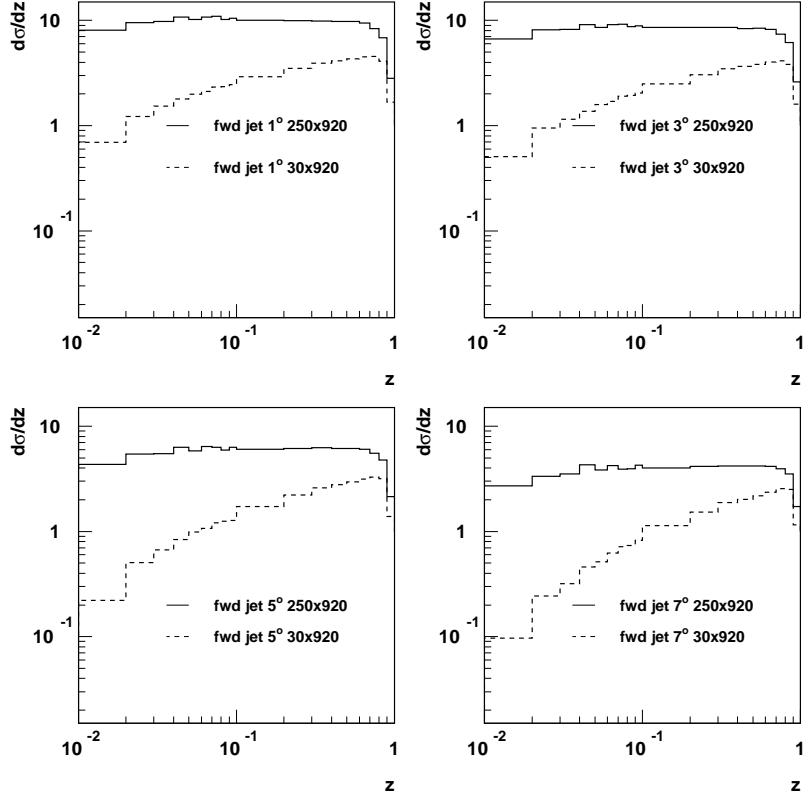,height=12.cm}
  \caption{{\it The values of the splitting variable $z$ for events satisfying the
  forward jet criteria, for different cuts on the minimum jet angle:
  $a.$  $\theta = 1\dg$, $b.$  $\theta = 3\dg$, 
  $c.$  $\theta = 5\dg$, $d.$  $\theta = 7\dg$. The solid line corresponds to the
  prediction at THERA energies, the dashed line corresponds to HERA energies.}
}
  \label{zval_fjet_thera} 
\end{center}
\end{figure}
In Fig.~\ref{zval_fjet_thera} we show the $z$ values obtained from
events satisfying the forward jet selection criteria at THERA energies
($\sqrt{s}=959$~GeV) compared with the ones at HERA energies
($\sqrt{s}=332$~GeV). The four plots correspond to different cuts on
the minimum jet angle ($\theta =1 \dg, 3 \dg, 5 \dg, 7 \dg$).  One
clearly can observe, that the distribution of $z$ values becomes flat
at small $z$ at THERA energies and that the small $z$ values are no
longer suppressed as in the HERA kinematic region. This clearly shows,
that the application of small $x$ evolution equations at THERA
energies are justified and unavoidable, although the sensitivity to
the treatment of large $z$ splittings does not completely go away.
\par
At HERA the forward jet cross section could be reasonably well
described by including a resolved virtual photon contribution, because
the phase space for small $z$ emissions at HERA energies was
relatively small. The situation changes at THERA energies: the phase
space for small $z$ emissions is larger and effects from small $x$
evolution become more visible. In Fig.~\ref{fig-fwdjet1} the cross
section for forward jet production is shown as a function of $x$ for
standard DGLAP prediction (dotted), including in addition a
contribution from resolved virtual photons (dashed-dotted) and the
CCFM prediction (dashed). For comparison the prediction from
ARIADNE\cite{ariadne}\footnote{using the tuned parameters given by
  `set2' in \cite{MCHERAtune}} is shown,
   which implements a semi-classical soft
radiation model in a dipole cascade, and which currently gives the
best overall description of small-$x$ HERA data. Since the the
available phase space at THERA is enlarged by a factor of $\sim10$,
the difference between the standard DGLAP-based calculation and the
CCFM calculation is increased.  Moreover, at THERA the CCFM approach
predicts a larger cross section than the model with resolved virtual
photon contributions added, and smaller cross section than ARIADNE,
while all three models give comparable results at HERA.  This gives a
unique opportunity, not only to distinguish between the different
approaches, but also to study details of the QCD cascade in a regime,
where the new small-$x$ evolution equations should be appropriate.

\begin{figure}[htb]
\begin{center}
\begin{minipage}[t]{0.49\textwidth}
  \begin{center}
   \hspace*{-1.5cm}
  \epsfig{file=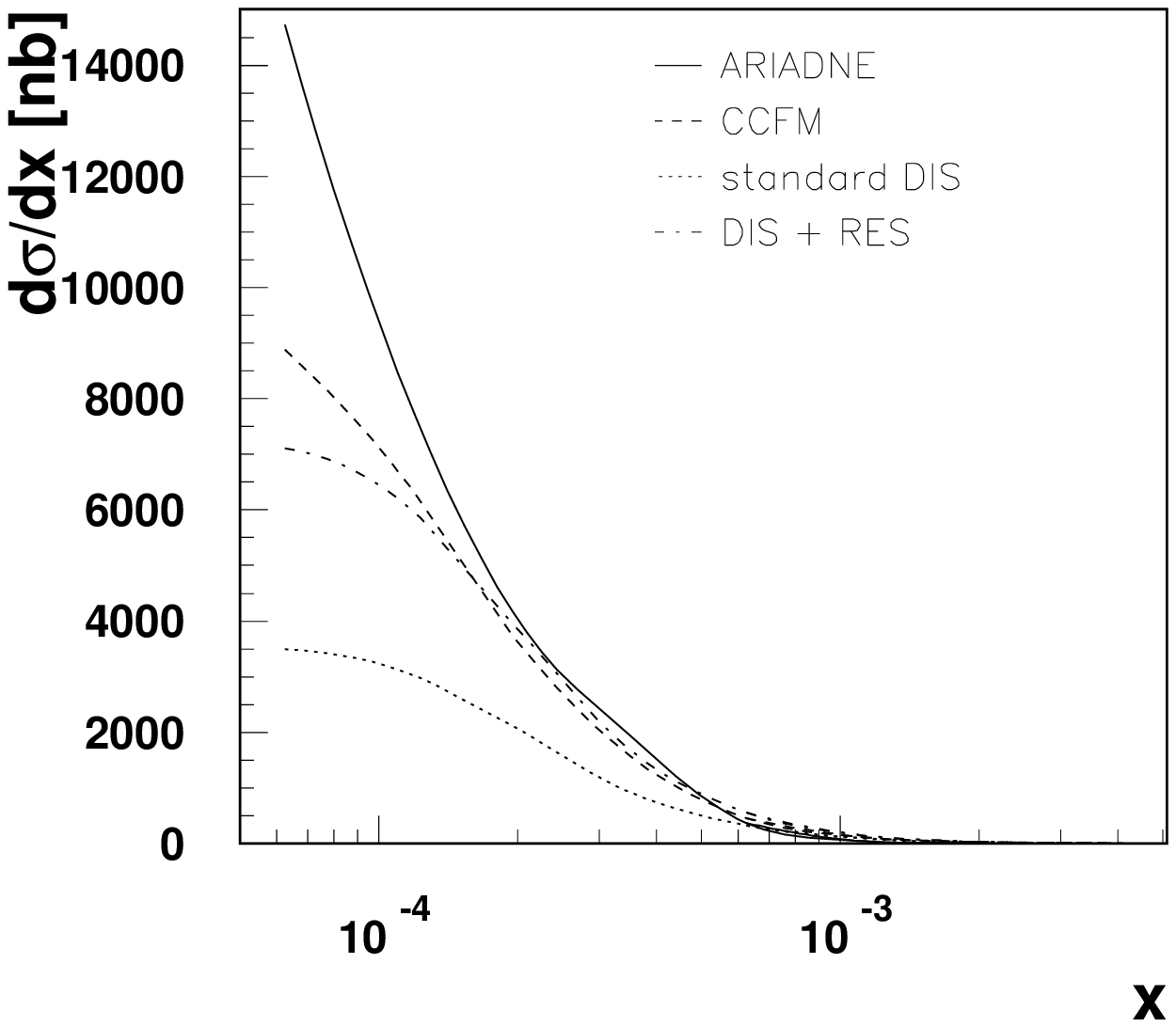,height=8.cm}
  \end{center}
   \vspace*{-1.5cm}
  \caption{{\it 
    Forward-jet cross section as a function of $x$ in different models for
    $0.5<p_t^2/Q^2<2$ and a minimum jet angle of $1\dg$}}
  \label{fig-fwdjet1} 
\end{minipage}\hfill
\begin{minipage}[t]{0.49\textwidth}
  \begin{center}
   \hspace*{-1.0cm}
  \epsfig{file=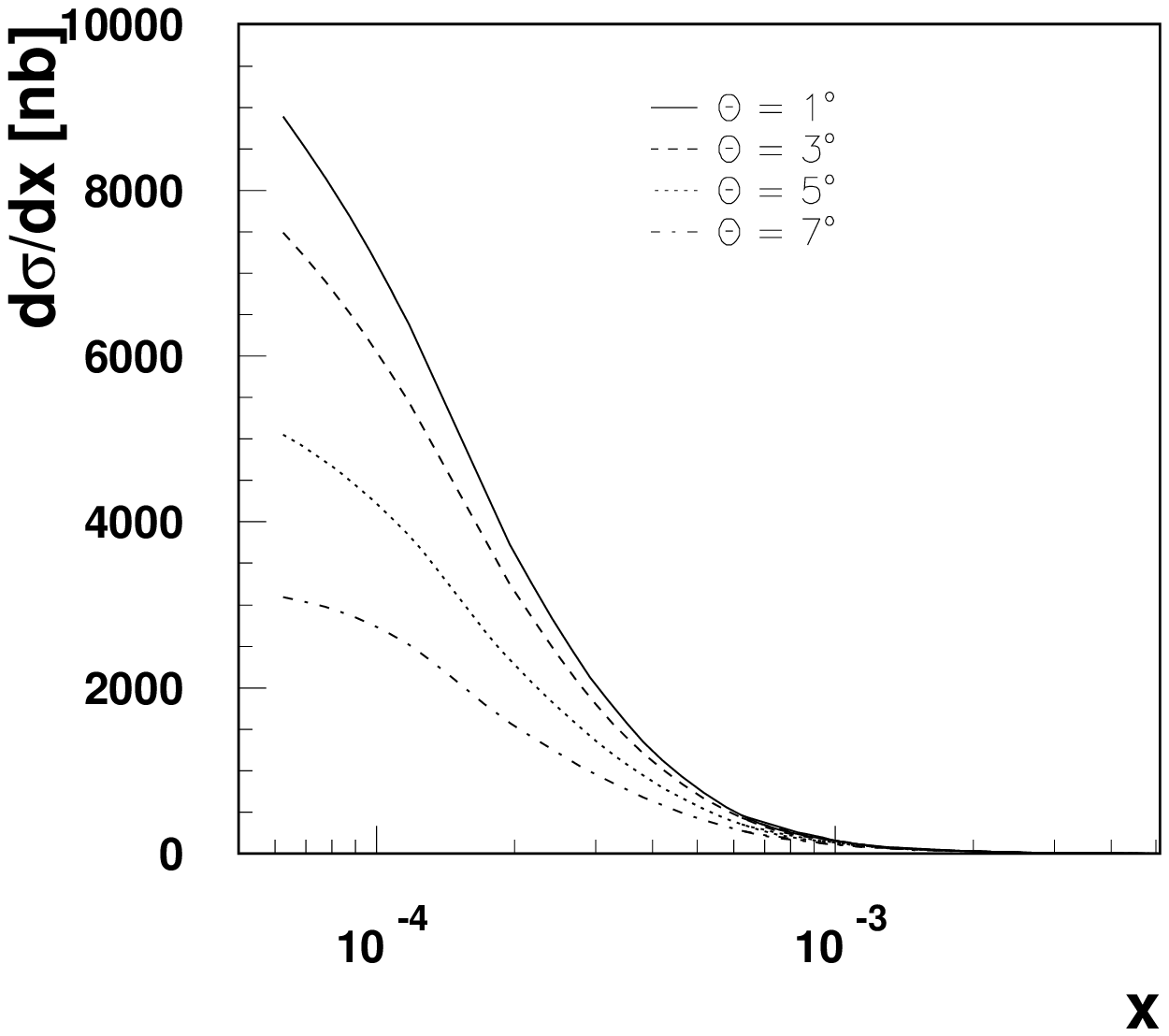,height=8.cm}
  \end{center}
   \vspace*{-1.5cm}
  \caption{{\it 
    Forward-jet cross section as a function of $x$ obtained from CCFM for
    $0.5<p_t^2/Q^2<2$, for different cuts on the minimum jet angle}}
  \label{fig-fwdjet2} 
\end{minipage}
\end{center}
\end{figure}

In Fig.~\ref{fig-fwdjet2} the forward jet cross section is shown for
different minimal jet angle cuts. On the experimental side this
requires complete acceptance both in the electron and proton direction
down to the lowest possible angles. From the size of the cross section
$d\sigma/dx$ one would like to reach at least $\theta\sim3\dg$ for the
forward jet measurement, but there are other reasons to ask even for
$\theta\sim1\dg$. On the other hand, the luminosities needed for such
measurements are moderate so that this question can be settled within
one year of running at THERA.

\section{Conclusion}

HERA has conclusively shown the need to go beyond the standard DGLAP
evolution equations in order to explain the data on small-$x$ final
states. The reach in $x$ is, however, not quite enough to really study
details of the small-$x$ evolution to be able to distinguish between
different approaches, such as the CCFM evolution, resolved virtual
photons and the dipole cascade model. With the increased kinematical
region available at THERA, this will become possible. The steeply
rising cross section as $x$ gets smaller means that such measurements
can be done even with moderate luminosity. On the other hand the
demands on the forward coverage of the detector is more critical ---
ideally it should be possible to measure jets down to an angle of
$1\dg$.

Once a good understanding of the small-$x$ evolution is obtained, it
should be possible to use the underlying $k_\perp$-factorization
theorem in BFKL/CCFM, where observables are described in terms of
process--dependent off-shell matrix element and universal
un-integrated parton densities, to make firm predictions of any other
small-$x$ measurement, just as normal DGLAP parton densities and matrix
elements are used today at large $Q^2$. THERA will be the only place
where these un-integrated parton densities can be measured.

\end{document}

%% file: ladder.pstex_t
\begin{picture}(0,0)%
\epsfig{file=ladder.pstex}%
\end{picture}%
\setlength{\unitlength}{4144sp}%
\begingroup\makeatletter\ifx\SetFigFont\undefined%
\gdef\SetFigFont#1#2#3#4#5{%
  \reset@font\fontsize{#1}{#2pt}%
  \fontfamily{#3}\fontseries{#4}\fontshape{#5}%
  \selectfont}%
\fi\endgroup%
\begin{picture}(2959,4027)(4083,-3841)
\put(5637,-826){\makebox(0,0)[rb]{\smash{\SetFigFont{12}{14.4}{\familydefault}{\mddefault}{\updefault}$y,Q^2$}}}
\put(5637,-1618){\makebox(0,0)[rb]{\smash{\SetFigFont{12}{14.4}{\familydefault}{\mddefault}{\updefault}$x_n,k_n$}}}
\put(5637,-3163){\makebox(0,0)[rb]{\smash{\SetFigFont{12}{14.4}{\familydefault}{\mddefault}{\updefault}$x_0,k_0$}}}
\put(5637,-2017){\makebox(0,0)[rb]{\smash{\SetFigFont{12}{14.4}{\familydefault}{\mddefault}{\updefault}$x_{n-1},k_{n-1}$}}}
\put(5637,-2366){\makebox(0,0)[rb]{\smash{\SetFigFont{12}{14.4}{\familydefault}{\mddefault}{\updefault}$x_{n-2},k_{n-2}$}}}
\put(5637,-2746){\makebox(0,0)[rb]{\smash{\SetFigFont{12}{14.4}{\familydefault}{\mddefault}{\updefault}$x_{n-3},k_{n-3}$}}}
\put(6631,-1868){\makebox(0,0)[lb]{\smash{\SetFigFont{12}{14.4}{\familydefault}{\mddefault}{\updefault}$p_n$}}}
\put(6631,-2235){\makebox(0,0)[lb]{\smash{\SetFigFont{12}{14.4}{\familydefault}{\mddefault}{\updefault}$p_{n-1}$}}}
\put(6631,-2615){\makebox(0,0)[lb]{\smash{\SetFigFont{12}{14.4}{\familydefault}{\mddefault}{\updefault}$p_{n-2}$}}}
\put(6631,-2995){\makebox(0,0)[lb]{\smash{\SetFigFont{12}{14.4}{\familydefault}{\mddefault}{\updefault}$p_{n-3}$}}}
\put(6792,-1351){\makebox(0,0)[lb]{\smash{\SetFigFont{12}{14.4}{\familydefault}{\mddefault}{\updefault}$\Xi$}}}
\put(5062,-3747){\makebox(0,0)[lb]{\smash{\SetFigFont{12}{14.4}{\familydefault}{\mddefault}{\updefault}$p$}}}
\end{picture}